\documentclass[reprint,nofootinbib,amsmath,amssymb,
                aps,pra]{revtex4-2}

\usepackage{graphicx}% Include figure files
\usepackage{dcolumn}% Align table columns on decimal point
\usepackage{bm}% bold math
\usepackage[T2A]{fontenc}
\usepackage{latexsym,amsmath,amssymb,amsbsy}
\usepackage{float}
\usepackage{xcolor}
\usepackage{subcaption}

\newcommand{\bra}[1]{\mathop{\langle{#1}\left.\hspace{-5 pt}\right|}\nolimits}
\newcommand{\ket}[1]{\mathop{\left|\hspace{-4 pt}\right.{#1}\hspace{-1 pt}\rangle}\nolimits}

\makeatother

\begin{document}

\title{Beable-guided measurement theory}

\author{A.\,M. \surname{Aleshin}}
\email{aleshinam@my.msu.ru}
\author{V.\,V. \surname{Nikitin}}
\email{riseway@proton.me}
\author{P.\,I. \surname{Pronin}}
\email{petr_pro_iv@mail.ru}
\affiliation{Department of Theoretical Physics,
Faculty of Physics, M.V. Lomonosov Moscow State University, Moscow 119991, Russia.}

\date{\today}

\begin{abstract}
In quantum mechanics, randomness is postulated as a separate axiom. De Broglie's theory allows one to reproduce quantum phenomena from completely deterministic formalism. But the question of the quantum randomness emergency in the de Broglie-Bohm theory needs special attention. In the work [G. Tastevin, F. Laloë, Comptes Rendus. Physique, 2021, 22, 1, pp. 99-116], it was shown that it arises as a result of the device microscopic state influence on the measurement result. In our work, we investigate the genesis of the quantum randomness in the de Broglie's theory in more details. Namely, we investigate the target system and the device behaviour in the decoherence process and model the measurement of canonical-conjugate observables. We propose a thought experiment which tests the opportunity of the information transition using beable-parameters violating the uncertainty relation. We show that in the measurement process, the strong stochastic fluctuations of beable parameters arise randomising the system in accordance with the uncertainty relation. Nevertheless, we find anomalous models of measurement in which these fluctuations can be neglected. These special models require further investigation.
\end{abstract}


\maketitle

\section{Introduction}\label{intro}
In the foundation of quantum theory, the randomness of measurement outcomes is postulated as result of an uncontrollable interaction between the classical device and the target quantum system. The necessity of separate axioms describing a measurement is known in literature as the measurement problem \cite{oxford_handbook, neumann, bohr_1, bohr_2, Frauchiger, collapse, men, bar}. Due to the von Neumann insolubility argument \cite{neumann}, we can not deduce the projective postulate from the unitary evolution within usual quantum mechanics axiomatics. Several interpretations of quantum mechanics provide different solutions of the problem using additional axioms. In this sense, they are an expansion of quantum mechanics. Taking into account the development of modern measurement devices and also rapidly evaluated quantum information theory \cite{quant_inf, entropy, jordi_1, jordi_2, markovich_1, markovich_2}, such theories can provide new approaches and predictions for the quantum system behaviour.

The work is devoted to  reconstruct the quantum randomness and the quantum contextuality\footnote{In accordance with the contextuality principle, measurement of observables in quantum theory cannot be considered as the preexisting values identifying.} from the deterministic de Broglie-Bohm formalism\cite{deB, bohm_1, bohm_2, bohm_3, str_1, str_2, nic_1, nic_2, Foo, val, kent, Flack_1, Flack_2, Philip} via the description of beables\footnote{"Beable" is an alternative term for "hidden variable" commonly used in literature (it is contrasted with the "observable"). We will use terms "beables" and "de Broglie coordinates" as synonyms.} behaviour in the simplest von Neumann measurement model \cite{neumann}.

These issues of contextuality in de Broglie-Bohm theory are active investigated \cite{context, bric, Holland}. In the work \cite{context}, it is shown that measurement result is depends not only on de Broglie states of target system but also on de Broglie device state. Taking into account the macroscopic large number of particles composing a real device, such dependence leads to an unpredictability of measurement results -- the quantum randomness. It provides the contextuality of the de Broglie-Bohm theory. An important consequence of the quantum contextuality is the absence of the direct connection between observables and beables. For example, in the works \cite{kurt,naun,heim}, it is shown that such direct connection of observation momentum with de Broglie velocity leads to Born's rule violation in the de Broglie-Bohm formalism.

Thus, these investigations convincingly show that the quantum randomness in the de Broglie-Bohm theory is a consequence of the chaotically influence of macroscopic device particles. However, to the best of our knowledge, the de Broglie particles dynamics in a decoherence process of a target system state during measurement has not been modeled yet. So, in the work \cite{context}, a guiding equation is considered already after a decoherence process. Nevertheless, as it will be shown this dynamics plays the key role for the randomness genesis: the decoherense process is accompanied by strong fluctuations of the device de Broglie particles, admitting the quantum randomness. Moreover, this randomization mechanism works even if we does not consider macroscopic properties of the measurement device.

The purpose of the present work is investigation of beable behaviour throughout the target system state decoherence in the measurement. In our model the measurement outcome is encoded in device beables. We study the encoding information scattering about the observable when measuring a canonically conjugate one. It is verified whether the arising stochastic fluctuations provide the Heisenberg uncertainty relation fulfillment \cite{hies}.

All calculations are made in nominal units and $\hbar=c=1$ is used. The goal of all the calculations is to understand fundamental question and not to make quantitative predictions.

\section{Description of measurement process in the de Broglie-Bohm theory}\label{Sec_measur}
The simplest formulation of the de Broglie-Bohm theory is in coordinate representation. In this formulation, the wave function is the physical field, guiding de Broglie particles. Wave function obeys the Schr\"{o}dinger equation
\begin{equation}
    i \frac{\partial \psi}{\partial t} = \hat{H} \psi.
\end{equation}

The dynamics of beable is determined by the guidance equation
\begin{equation}
    \frac{dx}{dt}=\frac{j}{\rho},
\end{equation}
where $x$ is the position of the beable, $j$ is the probability current density, $\rho=|\psi|^2$ is the probability density calculated using the wave function. We assume the system to be in the quantum equilibrium \cite{val}, i.e., an initial probability density of the de Brogile particles to be equal to $\rho=|\psi|^2$. In the de Broglie-Bohm theory, beable dynamics plays a role of alternative to wave packet reduction in other versions of quantum mechanics.

For simplicity, we suggest the device and target system to be in pure states. Then, we model the von Neumann measurement process \cite{neumann} -- an entanglement between the device state-vector and the target system state-vector
\begin{equation} \label{measure_gen}
    \hat{U}\ket{I}(\alpha_1 \ket{\psi_1}+\alpha_2 \ket{\psi_2}) = \alpha_1 \ket{I_1} \ket{\psi_1} + \alpha_2 \ket{I_2} \ket{\psi_2},
\end{equation}
here $\hat{U}$ is evolution operator, the target system state-vector $\ket{\psi}$ and the device state-vector $\ket{I}$ guide the target system and device beables, respectively. Due to the entanglement of wave functions, device needles and target system de Broglie particles turn out to be correlated and, thus, their trajectories are intertwined. Thus, the measurement result is encoded in the device de Broglie coordinates \cite{str_2}. The ensemble of such trajectories provides an opportunity to observe what actually happens to the target system and the device during the measurement process and therefore, separate the influence of the device on the target system and the target system on the device.

\subsection{Simple measurement model}
A simple model of the evolution operator $\hat{U}$ can be constructed by introducing the Hamiltonian $\hat{H_I}$ of the interaction between the device and the target system. For simplicity, the device has one degree of freedom and its coordinate is $r$, its momentum is $k$. At the same time, it is considered massive to model its classic properties -- the absence of the corresponding wave-function spreading. In accordance with von Neumann's measurement theory, the interaction Hamiltonian must commute with the operator of the measuring value $\hat{F}$ and depend on the momentum of the device needle so that the needle coordinate encodes the measurement result
\begin{equation} \label{measure_com}
    [\hat{F},\hat{H_I}]=0, \quad \quad \hat{H_I} = H_I\left(\hat{F}, \frac{\partial}{\partial r}\right),
\end{equation}
The simplest form of such Hamiltonian is
\begin{equation} \label{measure_H_I}
    \hat{H_I} = -i \lambda\hat{F} \frac{\partial}{\partial r},
\end{equation}
here $\lambda$ is a coupling constant.

An evolution due to this Hamiltonian makes the needle move in different ways depending on the observable values.

In the case of large coupling constant we can omit kinetic terms, and the measurement is projective. Solution of the Schr\"{o}dinger equation in this approximation has the form
\begin{gather}\label{eq_entang}
    \psi(q,r,t) = \int df F(f,q) I(r-\lambda f t),
\end{gather}
here $I(r)$ is the initial device wave-function, $F(f,q)$ is the eigenfunction of the $\hat{F}$ corresponding to eigenvalue $f$, $q$ and $r$ are the coordinates of the target system and device, respectively. The initial wave function of the device is chosen to be real in order to minimize the influence of the initial device motion on the measurement outcome.

\subsection{Decoherence process}
In the de Broglie-Bohm theory, decoherence is responsible for the separation of wave packets corresponding to different measurement outcomes, while the dynamics of de Broglie particles is responsible for the "choice" between these outcomes.

In the considered case, the density matrix of the target system has the form
\begin{gather}\label{eq_matrix}
    \rho_{\psi} = {Tr}_{I} \ket{\psi} \bra{\psi}=\nonumber\\
    \int \int df'df \ket{f'}\bra{f} \int dr I(r-\lambda f't)I(r-\lambda ft),
\end{gather}
here $\ket{f}$ is the target system eigenvector corresponding to the physical value $f$.

The non-diagonal elements of the density matrix are proportional to products ${I(r-\lambda f't)I(r-\lambda ft)}$. Thus, decoherence time $T$ is the time during which this product vanishes due to the space separation of device wave packages
\begin{equation}\label{eq_overloop}
    I\left(r-\lambda f T\right)I\left(r-\lambda f' T\right)=0.
\end{equation}

To make the numerical estimation, we assume that the device wave-function has a Gaussian form before measurement. Then, from 3 sigma-rule, we find the following estimation for the decoherence time
\begin{equation}\label{measure_time}
    T = \frac{6\sigma }{\lambda(f' - f)},
\end{equation}
where $\sigma$ is the characteristic size of the device wave packet.

As one can see, the decoherence occurs faster in the strong coupling regime and for highly distinguishable values of the target system physical quantities.

\section{Coordinate measurement}\label{Sec_coord}
Now, let us construct a model of the coordinate measurement. The simplest device-system interaction Hamiltonian has the form
\begin{equation}
    \hat{H_I} = -i \lambda q \frac{\partial}{\partial r}.
\end{equation}

The wave function in the strong coupling constant regime (the conditions for the kinetic term neglecting are considered in Appendix \ref{sec_appendix_coord_time}) is an entangled state of the device and the target system in the coordinate representation
\begin{equation}\label{supq}
    \psi = \int Q(q') \delta (q-q') R(r-\lambda q't)dq',
\end{equation}
here $Q(q')$ is the wave function of the target system, ${\delta (q-q')}$ is eigenfunction of the coordinate operator, $R(r)$ is the wave function of the device at the initial of time $t=0$.

Corresponding guidance equations take the form:
\begin{equation}\label{eq_v_r}
    v_r=\lambda q, \quad v_q = 0
\end{equation}

As we can see, the measurement result is encodeed by the beable device velocity.

However, there is an anomaly in this model because the contextuality principle is violated -- the measurement result describes the target system before the measurement. 
Indeed, in accordance with (\ref{eq_v_r}), the de Broglie device velocity is determined by the initial de Broglie coordinate of the target system. Moreover, the reduction is performed before the decoherence process is completed (see the formula (\ref{measure_time})). The contextuality violation provides a potential opportunity for the uncertainty principle violation discussed in the Sec. \ref{Sec_Hisenberg}.

The existence of these anomalous models show the importance of the decoherence process investigation that was skipped in the work \cite{context}. In Sec.~\ref{Sec_momentum}, we consider the momentum measurement model in which there is the restoring contextuality mechanism.

\section{The momentum measurement}\label{Sec_momentum}
To describe the momentum measurement, we used the simplest Hamiltonian of interaction with a device measuring momentum
\begin{equation}
    \hat{H_I} = -\lambda \frac{\partial}{\partial q}\frac{\partial}{\partial r}.
\end{equation}
Corresponding wave-function is an entangled state of the device and the target system in the coordinate representation
\begin{equation}\label{psi_p}
    \psi = \int dp' P(p')K(r-\lambda p't) \exp\left(i\left(p'q - \frac{{p'}^2}{2m}t\right)\right),
\end{equation}
here
$P(p')$ is wave function of the target system, $\exp\left(i\left(p'q\right)\right)$ is the eignfunction of the momentum operator,
$K(r-\lambda p't)$ is the wave function of the device.

Corresponding guidance equations are
\begin{gather}\label{eq_v}
    v_q = \frac{i\lambda}{2} \left(\frac{{\psi^*_r}'}{\psi^*} - \frac{{\psi_r}'}{\psi}\right) + \frac{i}{2m} \left(\frac{{\psi^*_q} '}{\psi^*} - \frac{{\psi_q}'}{\psi}\right),\\
    v_r = \frac{i\lambda}{2} \left(\frac{{\psi^* _q}'}{\psi^*} - \frac{{\psi_q}'}{\psi}\right).
\end{gather}

\begin{widetext}
Then we considered the superposition of $n$ momenta
\begin{equation}\label{eq_P_p1_pn}
	P(p')=A_1 \delta(p'-p_1 )+A_2 \delta(p'-p_2 )+...+A_n\delta(p'-p_n ).
\end{equation}

and obtained the following guidance equations for the target system and for the device
    \begin{gather}
    v_r = \frac{\lambda}{2} \left(p_1 \left(\frac{\psi_1}{\psi_1 + ... + \psi_n} + \frac{\psi_1^*}{\psi_1^* + ... + \psi_n^*}\right) + 
    ... + p_n\left(\frac{\psi_n}{\psi_1 + ... + \psi_n} + \frac{\psi_n^*}{\psi_1^* + ... + \psi_n^*}\right)\right) = \nonumber\\
    = \frac{\lambda}{2} \frac{\sum_{i=1}^n \sum_{j=1}^n A_i A_j K_i K_j (p_i + p_j)\cos{\alpha_{ij}}}{\sum_{i=1}^n\sum_{j=1}^n A_i A_j K_i K_j\cos{\alpha_{ij}}}, \label{v_r}\\
    v_q = \frac{v_r}{\lambda m} + \frac{\lambda}{2} \frac{\sum_{i=1}^n \sum_{j=1}^n A_i A_j K_i K_j\left(\frac{{K_j}'}{K_j} - \frac{{K_i}'}{K_i}\right)\sin{\alpha_{ij}}}{\sum_{j=1}^n \sum_{i=1}^n A_i A_j K_i K_j\cos{\alpha_{ij}}}\label{v_q},
    \end{gather}
    \begin{gather}\label{K_phases}
    K_i=K\left(r-\lambda p_i t\right), \quad
    {K_i}' = \frac{\partial K\left(r-\lambda p_i t\right)}{\partial r},
    \quad\alpha_{ij} = \left[(p_i-p_j)q + \frac{1}{2m} (p_j^2 - p_i^2) t\right].
    \end{gather}

From the equations, you can see that at the end of the measurement process, the product $K_i K_j$ vanishes, and the velocity of the device and the measured particle take the values one of the $p_i/m$ and $p_i$ values, respectively.

To describe the measurement process in details, guidance equations are solved numerically. On the Fig.~\ref{fig_vel}, the dependencies of the device needle velocity on time are shown under various initial conditions. The initial device wave-function is taken in the Gaussian form. The initial distribution for the device and the target system beables are selected in accordance with the Born rule.

\begin{figure}[H]
    \centering
    \begin{subfigure}[b]{0.4\textwidth}
        \includegraphics[width=\textwidth]{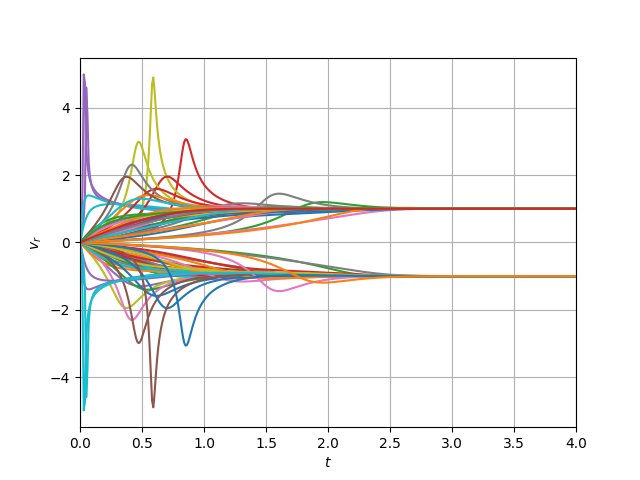}
        \caption{}
        \label{fig_vel_0.5}
    \end{subfigure}
    \begin{subfigure}[b]{0.4\textwidth}
        \includegraphics[width=\textwidth]{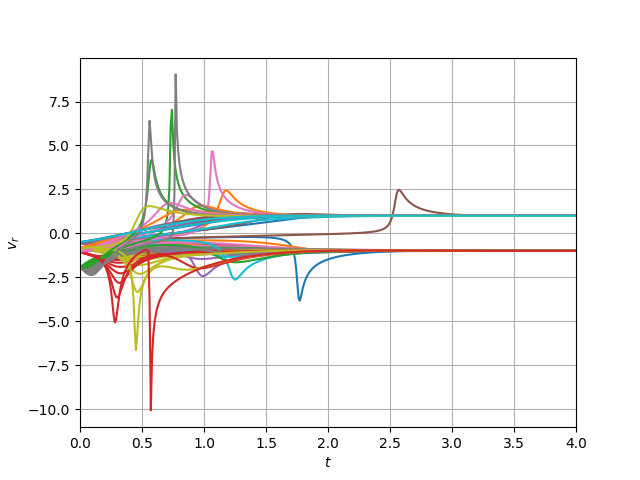}
        \caption{}
        \label{fig_vel_1/9}
    \end{subfigure}
    \begin{subfigure}[b]{0.4\textwidth}
        \includegraphics[width=\textwidth]{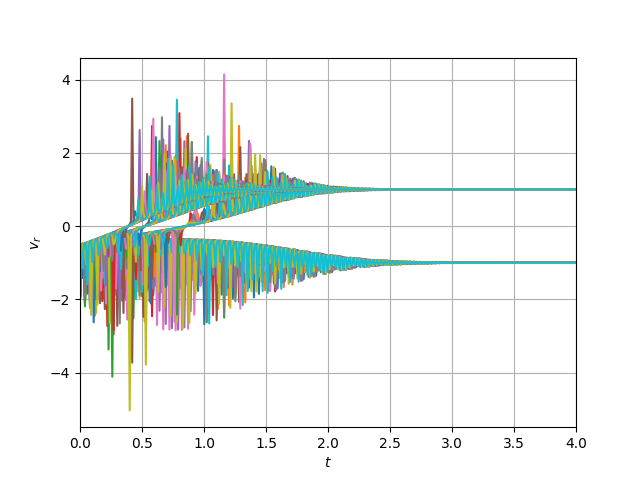}
        \caption{}
        \label{fig_vel_100}
    \end{subfigure}
    \caption{The dependence of the device needle velocity on time under different initial conditions for the momentum measurement case: the measurement of the target system stated in a superposition of two momenta $p_1 = 1, p_2 = -1$ with coupling constant $\lambda = 1$, particle mass $m = 1$, and different probabilities a) $A_2^2=A_1^2=0.5$, b) $A_1^2=1/10$, $A_2^2=9/10$, c) $A_1^2=1/10$, $A_2^2=9/10$, $m = 0.01$ (stochastic fluctuations increase with the target system mass decrease). Lines of different colors correspond to different initial positions of the target particle.}
    \label{fig_vel}
\end{figure}
\end{widetext}

The describing fluctuations occur in the decoherence process. For the examples on Fig.~\ref{fig_vel_0.5} and Fig.~\ref{fig_vel_1/9}, the measurement time estimation from the formula~(\ref{measure_time}) is $T=3$ ($p_1 = 1, p_2 = -1, \lambda = 1, \sigma = 1$). It can be seen, that most of the trajectories reach the stationary state before this time.

Therefore, we have the idea of the quantum randomness emergency of the measurement outcomes. During decoherence process, device de Broglie coordinates strongly fluctuate and randomly distributed by the end of the measurement. The measurement outcomes are determined by the device initial state - in full agreement with the contextuality principle and conclusions of the work \cite{context}. It distinguishes this model from the anomalous model of the coordinate measurement \ref{Sec_coord}. It will be shown in Sec.~\ref{Sec_Hisenberg}, the Heisenberg uncertainty relation is also directly related to these fluctuations.

\newpage
\subsection{Born's rule for the momentum probability distribution}\label{Sec_Born}
If we associate the velocity of a de Broglie particle with its momentum $p=m\frac{dx}{dt}$ as it was done in \cite{kurt,naun,heim}, then an incorrect probability distribution in the momentum representation will arise. However, now we show that the explicit simulation of the measurement process provides a correct momentum distribution.

\begin{figure}[H]
    \centering
    \includegraphics[scale=0.5]{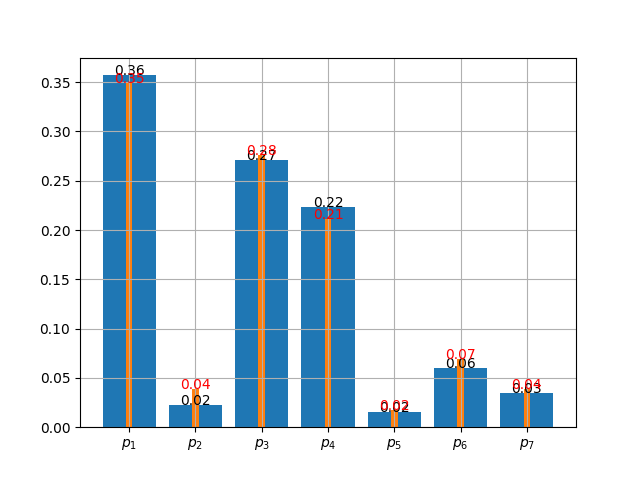}
    \caption{The momentum probability distribution obtained by the numerical simulation of the measurement at $A_1/A_2/A_3/A_4/A_5/A_6/A_7 = 9/3/8/7/2/4/3$. Wide columns -- the numerical result, thin lines -- theoretical predictions. The results are in a good agreement with the Born rule.}
    \label{fig_momentum_rusp}
\end{figure}

Having distributed the de Broglie device particles in accordance with Born's rule at the initial moment of time, we numerically find the distribution of the device de Broglie velocities after the measurement. In Fig. \ref{fig_momentum_rusp}, it is clearly seen that this distribution reproduces the distribution in the target system that was in a superposition state with $n=7$ momenta at a random probability amplitude.

To solve the problem, we refused to rely on the direct connection between the measured momentum and the de Broglie velocity. It means that the de Broglie velocity of the particle is not a quantity directly measured experimentally -- it is a hidden parameter. Moreover, the direct connection between the observable and the hidden parameter contradicts the quantum contextuality.

\subsection{Contextuality}\label{Sec_momentum_context}
\begin{figure}[H]
\centering
\includegraphics[scale=0.55]{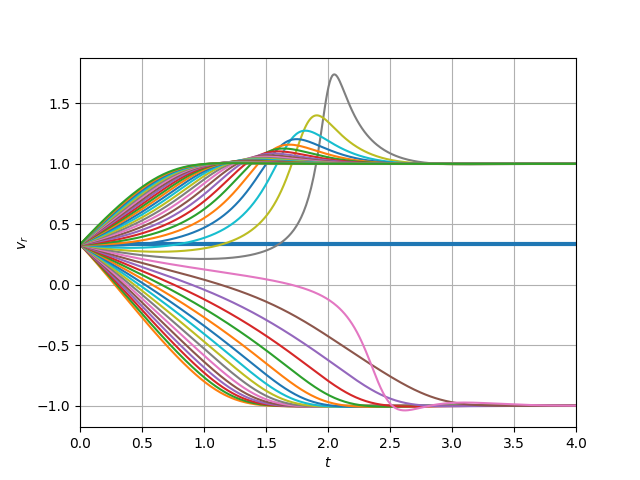}
\caption{The dependence of the device velocity on the time under different device initial conditions for the case of $p_1 = -1, p_2 = 1$, $A_1^2/A_2^2=1/4$, $\lambda = 1$, $m = 1$ with the fixing initial condition $q=0$ for the target system. Measurement outcomes depend on initial states of device particles. It is the quantum contextuality. In the classical limit (bold line) the contextuality is absent.}
\label{fig_inf}
\end{figure}

In the de Broglie-Bohm theory, the active influence \cite{bohr_2} of the device on the measurement result (quantum contextuality) is associated with the dependence of the measurement result on the initial states of the device particles \cite{bric, context}. The influence can be observed in the momentum measurement process. It is demonstrated on Fig.~\ref{fig_inf}, where initial conditions for the target system are fixed: as one can see, the measurement outcome does depend on the initial position of the device needle. 

\section{The uncertainty principle}\label{Sec_Hisenberg}

Since the de Broglie's theory is deterministic in its content, one may expect that during sequential measurements some information can be transferred using beable-parameters bypassing the uncertainty relation. Let us consider the following thought experiment. At first, we measure the coordinate of the particle then its momentum and then the coordinate again. After the first coordinate measurement, the wave packet is localized. After momentum measuring, the wave packet spreads (which causes the uncertainty ratio in quantum mechanics). However, the de Broglie particle may retain the information about the position of the particle at the localization moment. In order for the uncertainty relation to be fulfilled, it is necessary the sufficiently strong unpredictable changes are made to the de Broglie particle coordinate during the momentum measurement. Otherwise, when re-measuring the coordinate (after the momentum measuring), we can predict the measurement result with the greater accuracy than the axioms of quantum mechanics allow.

We use devices that are described in the previous sections for the verification of uncertainty relation. The Schr\"{o}dinger equation for this process
\begin{gather}\label{eq_sh_p_q}
    i \frac{\partial \psi}{\partial t} + i q \lambda_1 \frac{\partial \psi}{\partial r_1} \theta \left(T_1 - t\right) + \lambda_2 \frac{\partial^2 \psi}{\partial q \partial r_2}\Delta \theta_p + \nonumber\\
    + i q \lambda_3 \frac{\partial \psi}{\partial r_3} \Delta \theta_q + \frac{1}{2m} \frac{\partial^2\psi}{\partial q^2} = 0,
\end{gather}
here
\begin{gather}
    \Delta \theta_p = \theta \left(t - T_1\right) - \theta \left(t - T_1 - T_2 \right),\\
    \Delta \theta_q = \theta \left(t - T_1 - T_2 \right) - \theta \left(t - 2 T_1 - T_2 \right),
\end{gather}
$T_1, T_2$ are the measurement time determined by (\ref{measure_time}).

After the first coordinate measurement, the target system wave-function becomes superposition of localized states. Since the interaction with the first device is finished after the first measurement, only one of these locaized states influence the beable dynamics -- that which corresponds to the first measurement outcome. For our purposes this localized state can be approximated as:
\begin{gather}\label{eq_delta_fin}
\delta_R(q - q') = \sum_{n=-N}^{N} \exp\left(ip_n(q-q')\right),
\end{gather}
that is, after the first measurement we have in accordance with formula (\ref{eq_P_p1_pn})
\begin{gather}\label{eq_delta_fin}
P(p') = \sum_{n=-N}^{N} \delta (p' - \Delta p n).
\end{gather}
The beables dynamics in the second (momentum) measurement is described by guidance equations (\ref{v_r}), (\ref{v_q}).

\begin{figure}[H]
\centering
\includegraphics[scale=0.55]{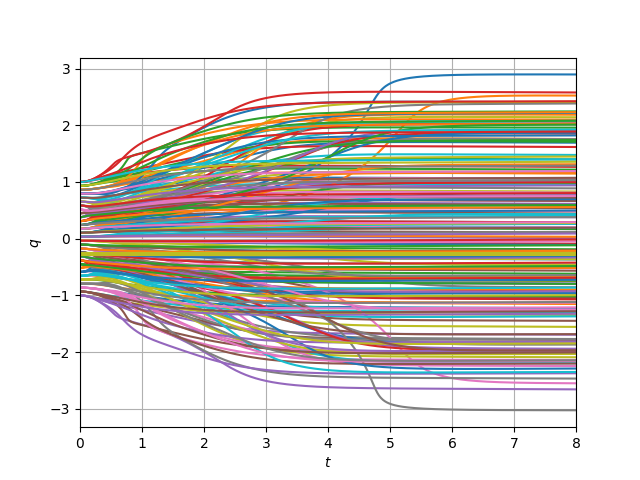}
\caption{Dependence of the target system de Broglie coordinates on time during momentum measurement. Different lines correspond to different initial positions of the particle. It is seen that target system beables, localized after coordinate measurement, are spread during the momentum measurement. Standard coordinate deviation increases from 0.21 to 1.19. Here, an initial time is $T_1 = 0$.}
\label{fig_q(t)_11}
\end{figure}

\begin{figure}[H]
\centering
\includegraphics[scale=0.55]{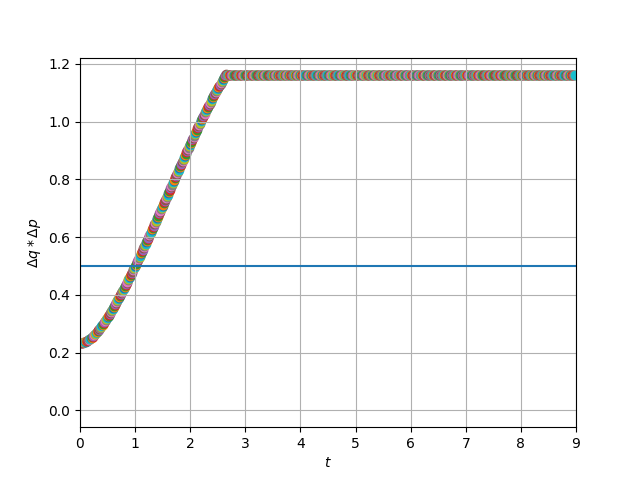}
\caption{The dependence of the product $\Delta p\Delta q$ on time during momentum measurement. $\Delta q$ is the coordinate dispersion of a particles group with a certain measured momentum, $\Delta p=1$ is the final uncertainty of the momentum after the measurement time. The thin line marks the lower limit $\frac{\hbar}{2} = 0.5$. Parameters: $m=1000$, $\lambda = 1$, $p_1=-5,...,p_{11}=5$ in conventional units.}
\label{fig_up_11}
\end{figure}

To check the uncertainty relation fulfillment, we estimate the final momentum standard deviation as $\Delta p$ (distance between delta-functions in approximate formula (\ref{eq_delta_fin}). To estimate standard deviation $\Delta q$, we assume that beables are distributed according to Born rule in the beginning of the momentum measurement and than evaluate them according to the guidence equations. In Fig.~\ref{fig_q(t)_11}, the results of the numerical solution for the target system de Broglie coordinates  are presented in the case of $2N+1=11$, $\Delta p = 1$ in formula (\ref{eq_delta_fin}). Fig. \ref{fig_up_11} shows the dependence of the product $\Delta p \Delta q$ on time. It can be seen from the figure that the stochastic perturbations arising in the measurement process are sufficient to fulfill the uncertainty relation.


Thus, stochastic fluctuations of the de Broglie particles coordinates introduced by the momentum measured device provide the mechanism for the restoring the uncertainty relation. These fluctuations spread the information about the coordinate encoded in the de Broglie particle.

However, the situation is not so clean in another thought experiment: at first momentum measurement, than the coordinate and the momentum once more. As it was shown in Sec.~\ref{Sec_coord}, the simple model admits a short-time coordinate measurements which induce no sufficient disturbances in the target system. Thus, we have a violation of the Heisenberg principle using this model. This issue is discussed in Sec.~\ref{sec_dis_conc}.

\section{Discussions and conclusions}\label{sec_dis_conc}
In the present work, de Broglie coordinates of the target system and measured device behaviour is investigated during decoherence process in the projective measurements of canonical-conjugate observables. For this purpose, the simplest measurement model of the coordinate and the momentum was developed.
The numerical modelling shows that the target system-measurement device interaction leads to strong stochastic fluctuations of the de Broglie velocities. These fluctuations provide the randomisation of measurement outcomes and the loss of the information about the canonical-conjugate observable in accordence with the uncertainty relation. This result confirms the conclusions of the work \cite{context}: the quantum contextuality in the de Broglie-Bohm theory is associated with the dependence of the measurement result on the initial states of the device particle.

Demonstrated stochastic fluctuations of the de Broglie device coordinates provide the Born's rule fulfillment for target system and device states after the measurement. In particular, the correct probability distribution in the momentum representation was restored. Thus, we provide the solution to the problem discussed in the works \cite{kurt,naun,heim}, where the direct connection between observable momentum and the de Broglie velocity was assumed.

Anomalous measurement models are found in which fluctuations of the device de Broglie coordinates are absent or are not enough intensively. In such models, spreading of the information about the obseravable in the strong measurement of canonical-conjugate observable is not sufficient to fulfill the uncertainty and contextuality principles. Such models provide the hypothetical possibility of the information encoding and the transportation using the de Broglie particles without corresponding losses in the strong measurement. Coordinate measurement model described in Sec. \ref{Sec_coord} is an example of the anomalous measurement model.

The consideration of the anomalous measurement models requires further research. Most probably, more realistic models of the device-target system interaction will be non-anomalous. However, there is a possibility that anomalous models open up new prospects for the experimental verification of the de Broglie-Bohm theory and practical applications. \\

\section*{Acknowledgements}

We would like to thank Yuri P. Rybakov, Anatoly V. Borisov, Kirill A. Kazakov, Oleg G. Kharlanov, for useful discussions. We also acknowledge the support of the Department of theoretical Physics, MSU. We thank Olga Kunitsyna and Sofia Kulikova for editing the English translation of our article.

\appendix
    \section{Coordinate measurement time estimation} \label{sec_appendix_coord_time}
To estimate conditions for the applicability of the projective measurement approximation, we substitute the suggested solution (\ref{supq}) in the full Schr\"{o}dinger equation
\begin{equation}
    i\frac{\partial \psi}{\partial t} + i \lambda q \frac{\partial \psi}{\partial r} + \frac{1}{2m} \frac{\partial^2\psi}{\partial q^2} = 0,
\end{equation}
and require that the kinetic term influence on the wave-function dynamics be small in comparison with the influence of interaction with the device.

\begin{equation}
    \frac{1}{2m} \left(\lambda^2 t^2 R'' Q - 2 \lambda t R' Q' + R Q''\right) \ll \lambda \Delta q R' Q,
\end{equation}
where $\Delta q$ -- the required measurement error (we estimate the relatively interaction influence on the wave function components with different coordinates), $R$ and $Q$ are device and target system wave functions, and strokes denote spatial derivatives.

In order for the decoherence to be fully determined by the projective approximation, it is necessary for the condition to be fulfilled throughout the entire process. Taking into account measurement time estimation (\ref{measure_time}) $T = \frac{6\sigma}{\lambda \Delta q}$ and estimating derivatives as $Q' \sim \Delta p Q$, $Q'' \sim \Delta p^2 Q$, $R' \sim \sigma R$, $R'' \sim \sigma^2 R$, we get the condition for the projective approximation:
\begin{equation}
    \lambda \gg \frac{\Delta p \sigma}{m\Delta q}.
\end{equation}
here $\Delta p$ is the characteristic momentum spread of the target system, $\sigma$ is the spatial width of the device wave function, $\Delta q$ is the measurement accuracy.

\end{document}